\documentclass[nofootinbib, prd,12pt,preprint,longbibliography,aps]{revtex4-1}
\usepackage{graphicx}
\usepackage{epstopdf}
\usepackage{amsmath}
\usepackage{amssymb}
\usepackage{amsfonts}
\usepackage{amsthm}
\usepackage[usenames]{color}
\usepackage{array}
\usepackage{verbatim}

\usepackage[
      colorlinks=true,
      linkcolor=blue,
      urlcolor=blue,
      filecolor=blue,
      citecolor=red,
      pdfstartview=FitV,
      pdftitle={},
      pdfauthor={},
      pdfsubject={},
      pdfkeywords={},
      pdfpagemode={},
      bookmarksopen=true
]{hyperref}

\usepackage{epsfig}

\usepackage{amsfonts,amsthm,amsmath,amssymb,graphicx}
\usepackage{hyperref}
\usepackage{amsmath}
\usepackage{url}
\usepackage{graphicx,color}

 \def\th{{\theta}}
 \def\a{{\alpha}}
 \def\frac#1#2{{#1\over #2}}

 \def\g{{\gamma}}

 \def\b{{\beta}}

\def\be{\begin{equation}}
\def\ee{\end{equation}}
\def\ba{\begin{eqnarray}}
\def\ea{\end{eqnarray}}
\numberwithin{equation}{section}

\begin{document}
\title{ Memory Effect and BMS-like Symmetries for Impulsive Gravitational Waves }
\author{Srijit Bhattacharjee \footnote{srijuster@gmail.com}}
\affiliation{${^a}$Indian Institute of Information Technology (IIIT), Allahabad,
Devghat, Jhalwa, Uttar Pradesh-211015, India}
\author{Arpan Bhattacharyya \footnote{bhattacharyya.arpan@yahoo.com} }
\affiliation{${^b}$Yukawa Institute for Theoretical Physics (YITP), Kyoto University,
Kitashirakawa Oiwakecho, Sakyo-ku, Kyoto 606-8502, Japan}
\author{Shailesh Kumar \footnote{shaileshkumar.1770@gmail.com }}
\affiliation{${^c}$Indian Institute of Information Technology (IIIT), Allahabad,
Devghat, Jhalwa, Uttar Pradesh-211015, India}
\preprint{YITP-19-43}

\date{\today}
\begin{abstract}
Cataclysmic astrophysical phenomena can produce impulsive gravitational waves that can possibly be detected by the advanced versions of present-day detectors in the future. Gluing of two spacetimes across a null surface produces impulsive gravitational waves (in the phraseology of Penrose \cite{Penrose}) having a Dirac Delta function type pulse profile along the surface. It is known that BMS-like symmetries appear as soldering freedom while we glue two spacetimes along a null surface. In this note, we study the effect of such impulsive gravitational waves on test particles (detectors) or geodesics. We show explicitly some measurable effects that depend on BMS-transformation parameters on timelike and null geodesics. BMS-like symmetry parameters carried by the gravitational wave leave some ``memory" on test geodesics upon passing through them. 

\end{abstract}

\maketitle


\section{Introduction}

The ``memory effect" for gravitational waves \cite{Zeldovich, Braginsky-Thorne, Christodolou, Thorne, 4, Bieri} has attracted considerable attention in recent times due to i) its theoretical connection with asymptotic symmetries and soft-theorems \cite{Strominger-Zbdv, Strominger}, and ii) possibility of its detection in the advanced detectors like aLIGO \cite{Lasky:2016knh} or in LISA \cite{Favata,Lasenby,1,2,3}. Gravitational memory effect is described as permanent displacement (or change in velocity) of a system of freely falling particles (idealised detectors), initially at relative rest, upon passing of a burst of gravitational radiation. In asymptotically flat spacetimes, in the far region, this change can be related to the diffeomorphisms that preserve the asymptotic structure, i.e. with the action of Bondi-van der Burg-Matzner-Sachs (BMS) group \cite{BMS}. Further, it has been shown that this displacement or change in metric components due to the passage of gravitational radiation is just the Fourier transform of the Ward identity satisfied by the infrared sector (soft) of quantum gravity's S-matrix, corresponding to the BMS like symmetries (like Supertranslations). These findings and Hawking-Perry-Strominger's \cite{Hawking:2016msc, HPS} proposal- that black holes may possess an infinite number of soft hairs corresponding to the spontaneously broken supertranslation symmetry, has generated a lot of activities towards finding the BMS like symmetries near the horizon of black holes. However, this has already been accomplished in various ways. BMS group and its extended version (including superrotation)\cite{Barnich:2009se} has been recovered by analyzing the diffeomorphisms that preserve the near horizon asymptotic structure of black holes \cite{Pinoetal, Pinoetal1}. Recently, it has been shown that the BMS symmetries can be recovered at any null hypersurface situated at some finite location of any spacetime \cite{Chandrasekaran:2018aop}. In the context of soldering of two spacetimes across a null hypersurface (like black hole horizons), BMS-like symmetries have been recovered in \cite{Blau, Bhattacharjee:2017gkh}. Now a natural question arises: Is it possible to detect some memory effect near the horizon of a black hole or at a null surface placed at some finite location of the spacetime just like what was found in the far region? In terms of the emergence of extended symmetries near the horizon of a black hole, BMS memory has been indicated in \cite{Donnay}. The purpose of this note is to study the gravitational memory effect at a finite location of spacetime and obtain some detectable features on test particles or geodesics.\par
Memory effect for plane gravitational waves has been extensively studied in recent times \cite{Zhang1, Zhang2}. The search for memory effect has also been extended to impulsive plane gravitational waves \cite{Zhang3}\footnote{See \cite{Stein} for clarifications on the calculation method adopted in \cite{Zhang3}.}. Memory effect for plane-fronted impulsive waves on null geodesic congruence (massless test particles) has been found in \cite{MO}, where BMS supertranslation type memory has been obtained. We extend these studies for timelike geodesics (test particles or detectors). We show both supertranslation and superrotation like memory effects, when a pulse of gravitational wave crosses these test particles. We also show that there are jumps in expansion and shear for a null congruence encountering \textit {impulsive gravitational waves} (IGW). \par
IGW is an artefact of Penrose's `cut and paste' method to glue two spacetimes \cite{Penrose}.  Mathematically speaking, these are spacetimes having a Riemann tensor containing a singular part proportional to Dirac delta function whose support is a null or light-like hypersurface. This kind of signal can be represented unambiguously as a pure gravitational wave part and a shell or matter part. Such impulsive signals are produced during violent astrophysical phenomena like supernova explosion or coalescence of black holes. Naturally studying the effect of such IGW on detectors (placed nearby the source) would be important from astrophysical perspective also. In this note, we consider null hypersurfaces that represent the history of the IGW, generated by gluing two space-times across the null surface and study the effect of these bursts of radiation on geodesics crossing it. As shown in \cite{Blau} and \cite{Bhattacharjee:2017gkh}, the gluing is not unique and one can construct infinitely many thin null-shells related to each other via BMS like transformations (on either side of the surface). Due to this, the geodesics crossing the IGW suffer a memory effect induced by BMS symmetries . \par
In section (\ref{sec2}), we review singular null-shells in General Relativity and show how BMS-like soldering freedom arises when one tries to glue two spacetimes across a null-hypersurface. In section (\ref{sec3}), we study the memory effect for IGW on timelike geodesics (detectors). We show how the relative position of test particles changes due to the passage of such a pulse signal. Next, we study the effect of IGW on null congruences. Here we calculate the jump in optical properties of geodesic congruence namely shear and expansion parameters due to the passage of IGW indicating memory effect. Finally, we conclude with the discussion of the outcomes and future objectives.

 \section{IGW and emergence of BMS invariance on null-shells} \label{sec2}

IGW are generated when one tries to glue two spacetimes across a null-hypersurface. To see this, let us consider two manifolds $\mathcal{M}_{+}$ and $\mathcal{M}_{-},$  with corresponding intrinsic metrics $g^{+}_{\mu\nu}(x_{+}^{\mu})$ and $g^{-}_{\mu\nu}(x_{-}^{\mu}).$ Throughout this note, we will work in four dimensions. We use Greek alphabets for spacetime indices. Latin lowercase and uppercase letters are used for hypersurface and codimension-1 surface quantities respectively. Suppose a common coordinate system $x^{\mu}$ that has overlaps with $x_{\pm}^{\mu}$,  is installed across the common null boundary $\Sigma$ of $\mathcal{M}_{\pm}$. We set $x^{\mu}|_{\Sigma}=\zeta^a.$ Let $e^{\mu}_a=\partial x^{\mu}/\partial \zeta^a$ are tangent vectors to $\Sigma$, and $n$ is normal vector to the hypersurface defined as $n^{\a}= g^{\a\b}\partial_{\b}\Phi(x)$, where the equation of $\Sigma$ is given by $\Phi=0$. The sign of this normal is taken so that it points to the future of $\Sigma$. We also have to define an auxiliary null vector $N^{\pm}$ to complete the basis such that \cite{BI,Poisson}, \be \label{auxn} N.N|_{\pm}=0,\,n. N=-1,\, N_{\mu}\cdot e^{\mu}_{A}=0.\ee    The junction condition in terms of common coordinates states: 
 \begin{align}
 \begin{split} 
\label{junc}
 [g_{ab}]=g^{+}_{ab}-g^{-}_{ab}=0, \end{split}\end{align}
together with,
\begin{align}
\begin{split}
[e^{\mu}_{a}]=[n^{\mu}]=[N^{\mu}]=0.
 \end{split}
 \end{align}
 In terms of common coordinates the metric reads
\be \label{junc3}
g_{\mu\nu}=g^{+}_{\mu\nu}\mathcal{H}(\Phi)+g^{-}_{\mu\nu}\mathcal{H}(-\Phi),
\ee
where $\mathcal{H}(\Phi)$ is the Heaviside step function. The Riemann tensor takes the following form upon using the junction condition (\ref{junc}),
\be \label{junc5}
R^{\alpha}{}_{\beta\gamma\delta}= R^{+\alpha}{}_{\beta\gamma\delta}\mathcal{H}(\Phi)+R^{-\alpha}{}_{\beta\gamma\delta}\mathcal{H}(-\Phi)+\delta(\Phi) Q^{\alpha}{}_{\beta\gamma\delta},
\ee
where $Q^{\alpha}{}_{\beta\gamma\delta}=-\Big([\Gamma^{\alpha}{}_{\beta\delta}]n_{\gamma}-[\Gamma^{\alpha}{}_{\beta\gamma}]n_{\delta}\Big).$ Clearly, any non-vanishing component of  $Q^{\alpha}{}_{\beta\gamma\delta}$ indicates existence of {\it impulsive gravitational wave} supported on the null hypersurface $\Sigma$. Now, one can show that the glued spacetime satisfies Einstein's equation with the following stress tensor \cite{BH, BH-ijmp,BI,Poisson}, 
\be \label{junc6}
T_{\alpha\beta}=T^{+}_{\alpha\beta}\mathcal{H}(\Phi)+T^{-}_{\alpha\beta}\mathcal{H}(\Phi)+S_{\alpha\beta}\delta (\Phi),
\ee
 with $S_{ab}=S_{\alpha\beta}e^{\alpha}_{a}e^{\beta}_{b}$ is denoted as the stress tensor of the null hypersurface. The stress tensor $S^{\a\b}$ can be expressed as \cite{BH, BH-ijmp,BI,Poisson},
 \be 
S^{\alpha\beta}=\mu n^{\alpha}n^{\beta}+j^{A}(n^{\alpha}e^{\beta}_{A}+e^{\alpha}_{A}n^{\beta})+p\,  \sigma^{AB}e^{\alpha}_{A}e^{\beta}_{B}, \label{ST}\ee
where  $\sigma_{AB}$ is the metric of the spatial slice of the surface of the null shell and $A,B$ denote the spatial indices of the null surface. $\mu, J^{A}$ and $p$ are surface energy density, surface current (anisotropic stress), and pressure of the shell respectively. Note that, the null-hypersurface has now become a shell with some null matter supported on it. We identify the intrinsic quantities of the shell as follows,  
\begin{align}
\begin{split} \label{junc8}
\mu=-\frac{1}{8 \pi}\sigma^{AB}[\mathcal{K}_{AB}],\, J^{A}=\frac{1}{8 \pi}\sigma^{AB}[\mathcal{K}_{VB}],\,
p=-\frac{1}{8 \pi}[\mathcal{K}_{VV}],\end{split}
\end{align}
where, 
\be \label{jumpmet}
\gamma_{ab}=N^{\alpha}[\partial_{\alpha} g_{ab}]=2[\mathcal{K}_{ab}],
\ee
and $\mathcal{K}_{ab}=e^{\alpha}_{a}e^{\beta}_{b} \nabla_{\alpha}N_{\beta}$ is the {\it `oblique' } extrinsic curvature.  
The jump in partial derivative of $g_{\mu\nu}$ is proportional to $\g_{\mu\nu}$, that contains all necessary information of the properties of the shell:
\be \label{gamma-p}
[\partial_{\lambda}g_{\mu\nu}]=\gamma_{\mu\nu}n_{\lambda}
.\ee

 Next, we briefly review how BMS symmetries emerge while one tries to solder two spacetimes across a null hypersurface. As depicted in \cite{Blau,Bhattacharjee:2017gkh}, if one tries to find possible coordinate transformations that preserve the induced metric on a null hypersurface across which two spacetimes are glued, BMS-like symmetry transformations emerge. This amounts to solving the Killing equation for $g_{ab},$ the induced metric on $\Sigma$ \cite{Blau}. 
\be \label{killing}
\mathcal{L}_{Z} g_{ab}=0.
\ee

 \begin{itemize}
\item {\bf Supertranslation like freedom on Killing horizon:} \\

We work in Kruskal coordinates $(U, V, x^A)$. Coordinates for the  horizon  $\Sigma$ are $\zeta^{a}=\{V,x^{A}\},$ where $V$ is the parameter along the hypersurface generating null congruences.  In this coordinate the normal to the surface becomes, 
$ \label{non1} n^{\alpha}=(\partial_{V})^{\alpha}.$ Let us first consider Schwarzschild horizon, 
\be \label{smet}
ds^2= -2 G(r) dU dV +r^2(U,V)(d\theta^2+\sin^2(\theta) d\phi^2),
\ee
where $G(r)=\frac{16\, m^3}{r} e^{-r/2m}, \,\, U V= -\Big(\frac{r}{2m}-1\Big) e^{r/2m}.$
The null shell is located at Killing horizon $U=0.$  The Killing equation (\ref{killing}) becomes,
\be \label{gentrans}
Z^{V}\partial_V g_{AB}+ Z^C\partial_C g_{AB}+\partial_A Z^C g_{BC}+\partial_B Z^C g_{AC}= 0,
\ee
leading $Z^V=F(V,\theta,\phi)$. Furthermore, if one also demands $\mathcal{L}_{Z} n^a=0$, then it gives,
\be \label{killing3}
\partial_{V} Z^{V}=0.
\ee
This will further restrict the form of $Z^V$ as
$
Z^{V}= T(\theta,\phi).
$
Clearly, $Z$ generates the following symmetry, \be \label{trans}
V\rightarrow  V+ T (\theta,\phi),
\ee
and this is strikingly similar to supertranslation found at null infinity of asymptotically flat spacetimes \cite{BMS}. This construction has been extended for the rotating blackhole spacetimes in \cite{Bhattacharjee:2017gkh}.
\item {\bf Superrotation like soldering freedom:} 

If the non-degenerate subspace of a null surface is parametrized by some null parameters (eg. null coordinates $U$ or $V$) then (\ref{gentrans}) offers a new kind of solution. This has been shown in  \cite{Bhattacharjee:2017gkh}. This situation gives rise superrotations or conformal rescaling of the base space of null hypersurface. To see this, consider the following metric \cite{Bhattacharjee:2017gkh}:
\be \label{constcurvature}
ds^2=\frac{2 (\frac{V}{1+\eta\,  \zeta\,\bar \zeta})^2 d\zeta d\bar \zeta +2 dU \, dV-2\,\eta\, dU^2}{(1+\frac{\Lambda\, U}{6}(V-\eta\, U))^2} .
\ee
This metric can represent de Sitter space when $\Lambda >0,\eta=1$, anti-de Sitter space when $\Lambda <0,\eta=-1$ and Minkowski space when $\Lambda=0,\eta=0.$  The null surface is at  $U=0.$  For this case, the following soldering freedoms are obtained \cite{Bhattacharjee:2017gkh}
\begin{align}
\begin{split} \label{trans1}
&V\rightarrow V(1-\frac{\epsilon\,\tilde \Omega(\zeta,\bar \zeta)}{2}),\\&
\zeta\rightarrow \zeta+\epsilon\, h(\zeta), \bar \zeta \rightarrow \bar \zeta+ \epsilon\, \bar h(\bar \zeta),
\end{split}
\end{align}
where,
\be \label{Omega}
\tilde \Omega(\zeta,\bar \zeta)=\frac{(1+\eta\, \zeta \bar \zeta) \left(h'(\zeta)+\bar h'(\bar \zeta)\right)-2 \,\eta \, \bar \zeta h(\zeta)-2 \eta \, \zeta\, \bar h(\bar \zeta)}{1+\eta \, \zeta \bar \zeta}.
\ee
$h(\zeta)$ and $\bar h(\bar \zeta)$ are holomorphic and anti-holomorphic functions and $\epsilon$ is a small parameter. This kind of local conformal transformations are usually referred as superrotation in the context of asymptotic symmetries. Emergence of such symmetries in the context of Penrose's impulsive wave spacetime (introducing snapping cosmic string) has also been discussed in  \cite{Strominger:2016wns}. 

\item {\bf Null surface near Killing Horizon:}  

Superrotations can also be recovered  when a null surface situated just a little away from the horizon of a black hole \cite{Bhattacharjee:2017gkh}. Using similar type coordinates as in Eq.(\ref{constcurvature}) the 2-dimensional spatial slice of Schwarzschild black hole becomes $r^2 \frac{d\zeta d\bar \zeta}{(1+\zeta\,\bar \zeta)^2}$, with $r^2= 4m^2+ \left(-\frac{8 m^2}{e}\right)U\, V+ \mathcal{O} (U\, V)^2+ \cdots.$ A null shell located at $U=\epsilon$ allows following soldering transformations:
\begin{align}
\begin{split} \label{trans2}
&V\rightarrow V(1-\tilde \Omega(\zeta,\bar \zeta) )-\frac{a\,\tilde \Omega(\zeta,\bar \zeta) }{b \,\epsilon} +\mathcal{O}(\epsilon),\\&
\zeta\rightarrow \zeta+ h(\zeta),\,\, \bar \zeta \rightarrow \bar \zeta+ \bar h (\zeta).
\end{split}
\end{align}
\end{itemize}
Hence, when we glue a supertranslated (or superrotated) metric with a seed metric (eg. Schwarzschild), these BMS-like symmetries emerge in the shell's intrinsic quantities as well as in IGW part (see Appendix A). 

\section{Memory Effect: Timelike congruence} \label{sec3}

In this section, we study interaction between IGW and test particles following timelike geodesics.  We must find some observable effects on the relative motion of neighbouring test particles. To see this, we consider two test particles whose worldlines are passing through a null hypersurface supporting IGW. The effect of IGW on the separation vector of these nearby geodesics can be captured by the use of geodesic deviation equation (GDE) and junction conditions. Theory of impulsive gravitational waves is a well-studied area of research \cite{BH, Balasin, Steinbauer, Podolosky}. Usually, two different approaches are adopted. In \cite{ Zhang3,Balasin, Steinbauer, Podolosky} etc, a distributional metric is used and no null-shell is introduced.  Here, we use a local coordinate system in which the metric tensor is continuous but its first derivative has discontinuity across the null surface. As a result, we get a thin null-shell along with the wave signal. It should be noted, the treatment presented here is based on the fact that a consistent $C^0$ matching of the metrics can be done maintaining $C^1$ regularity of the geodesics across the null shells \cite{Steinbauer:2013spa}. Rigorous mathematical treatment of IGW spacetimes addressing the existence, uniqueness and generic properties of geodesics can be found in \cite{Podolsky:2014ysa, Podolsky:2019lvp, Podolsky:2016mqg,Podolsky:2010xh}. Here we closely follow the construction given in \cite{BH, BH-ijmp} and review the necessary parts.\par
Let us consider a congruence having  $T^{\mu}$ to be the tangent vector, where $T^{\mu}$ is a unit time-like vector field.
\begin{equation}
g_{\mu\nu}T^{\mu}T^{\nu} = -1
\end{equation} 
The integral curves of $T^{\mu}$ happens to be time-like geodesics. Therefore,
\begin{equation}
\dot{T}^{\mu} = T^{\nu} \nabla_{\nu}T^{\mu}= 0.
\end{equation}
Dot denotes the covariant derivative of a tensor field in $T^{\mu}$ direction. This congruence is passing through the null shell located at $\Sigma.$ Now we introduce a separation vector $X^{\mu}$ satisfying $g_{\mu\nu}T^{\mu}X^{\nu}=0$ such that,
\begin{equation}
\dot{X}^{\mu} =X^{\nu}\nabla_{\nu} T^{\mu} .
\end{equation} 
This vector $X^{\mu}$ satisfies the geodesic deviation equation,
\begin{equation}
\ddot{X}^{\mu} = -R^{\mu}_{\lambda\sigma\rho}T^{\lambda}X^{\sigma}T^{\rho}.
\end{equation}
$R^{\mu}_{\lambda\sigma\rho}$ is Riemann tensor of full space-time $\mathcal{M}$. To be consistent with the jump in transverse derivative of the metric (\ref{jumpmet}), which produces a Delta function term in Riemann tensor, we can assume the jumps in partial derivatives of $T^{\mu}$ and $X^{\mu}$ across $\Sigma$, and given by,
\begin{equation} \label{t1}
[T^{\mu}_{,\lambda}] \footnote{`,' means derivative.}= P^{\mu}n_{\lambda};\,\,\, [X^{\mu}_{,\lambda}] = W^{\mu}n_{\lambda},
\end{equation}
for some vectors $\mathcal{P}$ and $W$ defined on $\Sigma.$ The vectors $P, \,W$ are not necessarily tangential to $\Sigma.$
Let $\lbrace E_{a}\rbrace$ be a triad of vector fields defined along the time-like geodesics tangent to $T^{\mu}$ by parallel transporting $\lbrace e_{a}\rbrace$ along these geodesics. Therefore,
\begin{equation}
\dot{E}^{\mu}_{a} = 0,
\end{equation}
where, $E_{a} \big|_{\Sigma}=e_{a}$. Consequently the jump in the partial derivatives of $E^{\mu}_{a}$ should take the following form for some $F^{\mu}_{a}$ defined on $\Sigma$,
\begin{equation}
[E^{\mu}_{a,\lambda}] = F^{\mu}_{a}n_{\lambda}.
\end{equation}
Let $X^{\mu}_{(0)}$ is the vector $X^{\mu}$ evaluated on $\Sigma$ i.e. the value of separation vector $X^{\mu}$ at the intersection point of the null hypersurface with the timelike congruence. Similarly, let $T^{\mu}_{(0)}$ denotes the tangent vector $T^{\mu}$ evaluated on $\Sigma$. We can write $X^{\mu}_{(0)}$ in terms of components along the orthogonal and the tangential vectors to the hypersurface in the following way, \begin{equation} \label{t2}
X^{\mu}_{(0)} = X_{(0)}T^{\mu}_{(0)}+X^{a}_{(0)}e^{\mu}_{a},
\end{equation}
for some function $X_{(0)}$, and $X^{a}_{(0)}$ is the $X^{a}$ evaluated at $\Sigma$. Now if $\Sigma$ is located at $U=0$ (this happens typically in Kruskal-like coordinates) , then using (\ref{t1}), and expanding around $U=0$ (keeping only the linear term)  we get, 
\begin{equation} \label{t3}
X^{\mu} = X^{-\mu} +U\, {\cal{H}} (U)W^{\mu},
\end{equation}
with $X^{-\mu}$ in general having dependence on $U$ (in the $-$ side) such that when $U=0$, $X^{-\mu}=X^{\mu}_{0}$. It is convenient to calculate $X^{\mu}$ in the basis $\lbrace E^{\mu}_{a},T^{\mu}\rbrace$. Its non-vanishing components,
\begin{equation}
X_{a} = g_{\mu\nu}X^{\mu}E^{\nu}_{a}.
\end{equation} 
Using (\ref{t2}) and (\ref{t3}), we calculate the above expression for small $U>0$ (in $+$ side),
\begin{equation} \label{t6}
X_{a} = \Big(\tilde{g}_{ab}+\frac{1}{2}\,U\, \gamma_{ab}\Big)X^{b}_{(0)}+U\,V_{(0)a}^{-}
\end{equation}
where $V_{(0)a}^{-}=\frac{dX_{a}^{-}}{du} \big |_{U=0}$ and $\tilde{g}_{ab}$  given by (using the definition of $\gamma_{ab}$ in (\ref{jumpmet})),
\begin{equation} \label{t7}
\tilde{g}_{ab} = g_{ab} + (T_{(0)\mu}e^{\mu}_{a})(T_{(0)\nu}e^{\nu}_{b}).
\end{equation}
The last term in Eq. (\ref{t6}) denotes the relative displacement of test particles for small $U$ when no signal were present. 
As the $\tilde{g}_{ab}$ is non-degenerate we can invert it,
\begin{equation}\label{tt6}
X_{a} = \tilde{g}_{ab}X^{b},\qquad X^a=\tilde{g}^{ab}X_b.
\end{equation}
To see the effect of wave and shell part of the signal separately we decompose $\gamma_{ab}$ into a transverse-traceless part, and rest as:
\begin{equation}
\gamma_{ab} = \hat{\gamma}_{ab} + \bar{\gamma}_{ab}
\end{equation}
with,
\begin{equation} \label{barmet}
\bar{\gamma}_{ab} = 16\pi\, \Big(g_{ac}S^{cd}N_{d}N_{b}+g_{bc}S^{cd}N_{d}N_{a}-\frac{1}{2}g_{cd}S^{cd}N_{a}N_{b}-\frac{1}{2}g_{ab}S^{cd}N_{c}N_{d}\Big),
\end{equation}
where $S^{ab}$ is defined in (\ref{ST}) and we have also used the definitions (\ref{junc8}). 
Here
\begin{equation}\label{hgamma}
\hat{\gamma}_{ab} = \gamma_{ab}-\frac{1}{2}g_{*}^{cd}\gamma_{cd}g_{ab}-2 n^{d}\gamma_{d(a}N_{b)}+\gamma^{\dagger}N_{a}N_{b},
\end{equation}
and $g_{*}^{ab}$ is the \textit{pseudo-inverse of $g_{ab}$ with entries at the V-direction are zero} defined as \cite{BI},
\begin{equation}
g_{*}^{ab}g_{bc} = \delta^{a}_{c}-n^{a}N_{c},\,\, g_{*}^{cd}\gamma_{cd}= g^{AB}\gamma_{AB}.	
\end{equation}
Also, $\gamma^{\dagger}=\gamma_{ab}n^{a}n^{b}$ and $\gamma_{ab}$ is defined via (\ref{jumpmet}).
$\hat{\gamma}_{ab}$ encodes the pure gravitational wave degrees of freedom as $\hat{\gamma}_{ab}n^b=0=g_*^{ab}\hat{\gamma}_{ab}$. On the other hand, $\bar \gamma_{ab}$ encodes the degrees of freedom corresponding to the null matter of the shell. If $\bar \gamma_{ab}$ vanishes then we will have a pure impulsive gravitational wave. Otherwise, the IGW will be accompanied by some null-matter. \par
Next, from (\ref{barmet}) and using (\ref{junc8}) for Kruskal type coordinates we get,
\begin{align} \label{t5}
\bar \gamma_{VB}=\bar \gamma_{BV}= 16\pi g_{B C}S^{VC}, \bar{\gamma}_{AB} =& -8\pi\,S^{VV}g_{AB}.
\end{align}
Using (\ref{tt6}), (\ref{t7}), (\ref{barmet}) and (\ref{hgamma})  in (\ref{t6}) we get,
\begin{align}
\begin{split}
&X_{V} = X_{(0) V}+\frac{1}{2}\, U\, \bar{\gamma}_{VB}X^{B}_{(0)}+U\, V_{(0) V}^{-}, \\&
 X_{A} = g_{AB}X^{B}_{(0)}+\frac{1}{2}\, U\, \gamma_{AB}X^{B}_{(0)}+U\, V_{(0)A}^{-}
\end{split}
\end{align}
We assume before the arrival of signal the test particles reside at the spacelike 2-dimensional surface (signal front). Therefore, we must have, 
\be
X_{(0) V}= V_{(0)V}^{-}=0,
\ee
leading to (using (\ref{t5})),
\begin{align}
\begin{split}
&X_{V} = \frac{1}{2}\, U\, \bar{\gamma}_{VB}X^{B}_{(0)}= 8\pi \, U\, g_{B C}S^{VC} X^{B}_{(0)},\\&
 X_{A} = (1-4\pi\,U\,S^{VV})(g_{AB}+\frac{U}{2}\, \hat \gamma_{AB}\, ) X^{B}_{(0)}+U\, V_{(0)A}^{-}.
\label{dev}\end{split}
\end{align}
Note that, if the surface current $S^{V C}\neq0$, then $X_{V}\neq0,$ which indicates that test particles will be displaced off the 2-dimensional surface. Since $\hat{\gamma}$ does not affect $X^V$, this component of the deviation vector can only have non-vanishing change if the IGW becomes the history of a null-shell containing current. Note that, for plane fronted wave we can have shell without any matter, in which this part will be zero \cite{BH}. However, for spherical-shells (such as horizon-shell in Schwarzschild spacetime), we can't have a pure gravitational wave without matter \cite{Blau, Bhattacharjee:2017gkh}. For shells having  $S^{VC}=0$, we only have the spatial parts of the deviation vector non-vanishing,
\begin{align}
\begin{split}
X_{A} =  (1-4\pi\,U\,S^{VV})(g_{AB}+\frac{U}{2}\, \hat \gamma_{AB}\, ) X^{B}_{(0)}.
\label{dev1}\end{split}
\end{align}
The terms involving $\hat{\gamma}_{AB}$ in  (\ref{dev1}) describe the usual distortion effect of the wave part of the signal on the test particles in the signal front. Whereas there is an overall diminution factor in the first bracket of the above expression due to the presence of the null-shell. We now examine the expression of deviation vector components for different cases.
 
\subsection{ Memory effect and supertranslations at the black hole horizon}
We first consider horizon shell of Schwarzschild metric (\ref{smet}). For this case, following \cite{Blau} we have, 
 \begin{align}
 \begin{split}
\hat{\gamma}_{\th\phi} = 2\frac{\nabla_{\th}^{(2)}\partial_{\phi} F}{F_V},
\hat{\gamma}_{\th\th} = \frac{1}{2}\left(\gamma_{\th\th}-\frac{1}{\sin^2\th}\gamma_{\phi\phi}\right)
\end{split}
\end{align}
If we focus on the case  (\ref{trans}) for which we get supertranslation from the  horizon soldering freedom, we get a shell having zero pressure and zero surface current, i.e. $S^{VA}=0=p$ \cite{Blau}.
Then we get,
\be
\hat{\gamma}_{\th\phi}=2\nabla_{\th}^{(2)}\partial_{\phi}T (\theta,\phi),
 \hat{\gamma}_{\th\th}=2\left(\nabla_{\th}^{(2)}\partial_{\th}T (\theta,\phi)-\frac{1}{\sin^2\th}\nabla_{\phi}^{(2)}\partial_{\phi}T(\theta,\phi)\right).
 \ee
 Also the energy density becomes,
 \be
 S^{VV}=-\frac{1}{32 m^2\,\pi}(\nabla^2 T(\theta,\phi)-T(\theta,\phi)).
 \ee
This shell  allows impulsive gravitational waves along with some matter and the neighbouring test particles will encode this into the $X^A$ components of the deviation vector (\ref{dev}) ($X_{V}$ is zero):
\begin{align}
\begin{split}
X_{\th}=\Big(1+\frac{U}{8 m^2}\, (\nabla^2 T(\theta,\phi)-T(\theta,\phi))\Big)&\Big[\Big(4 m^2+U\,(\nabla_{\th}^{(2)}\partial_{\th}T (\theta,\phi)-\frac{1}{\sin^2\th}\nabla_{\phi}^{(2)}\partial_{\phi}T(\theta,\phi))\Big)X^{\theta}_{(0)}\\&+U\, \nabla_{\th}^{(2)}\partial_{\phi}T (\theta,\phi) X^{\phi}_{(0)}
\Big]
\end{split}
\end{align}
and similarly for $X^{\phi}$ component. 

Clearly, there is a distortion in the relative position of the particles after encountering the impulsive gravitational wave and the distortions are determined by the supertranslation parameter $T(\th,\phi)$. The displacement is confined to 2-surface and for physical matter (See (\ref{mu}) of the appendix) having a positive energy density,  there will be a diminishing effect on the test particles.  Hence, this is a reminiscence of BMS memory effect (velocity) at future null infinity. We may integrate this expression with respect to the parameter of the geodesics to get a shift in the spatial direction and obtain the displacement memory effect. For a slowly rotating spacetime, we can get similar kind of memory effect following \cite{Bhattacharjee:2017gkh}.

\subsection{ Memory effect and superrotation like transformations}

To see the effect of superrotation on test particles, we first focus on `spherical' impulsive wave signal introduced by Penrose. The analysis for hyperboloidal wave can be done similarly.  Extending the transformation (\ref{trans1}) off the null shell, we can compute the $\hat \gamma_{AB}$ and the stress tensor (\ref{ST}). The details of the computation is given in the appendix A. We simply quote the important results below.  
\be
\hat \gamma_{\zeta\zeta}= \epsilon\, V\, h'''(\zeta), \hat \gamma_{\bar \zeta \bar \zeta}=\epsilon\, V\, \bar h'''(\bar \zeta).
\ee
Also we get,
\be
S^{VV}=0, S^{V\zeta}=j^{\zeta}=\frac{\epsilon\, \bar h''(\bar \zeta)}{16\pi\, V^2}, S^{V\bar \zeta}=j^{\bar \zeta}=\frac{\epsilon\,  h''( \zeta)}{16\pi\, V^2}.
\ee
Using these we get,

\begin{align}
\begin{split}  \label{restf}
X_{\zeta} = V\Big(V\, X^{\bar \zeta}_{(0)}+ \epsilon\,\frac{U}{2}\, h'''(\zeta) X^{\zeta}_{(0)}\Big) ,X_{\bar \zeta } =  V\Big(V\, X^{\zeta}_{(0)}+\epsilon\, \frac{U}{2}\,  \bar h'''(\bar \zeta) X^{\bar \zeta}_{(0)}\Big).
\end{split}
\end{align}

Here $\epsilon$ is the infinitesimal parameter of conformal transformations. Also note that, unlike the previous case we have some non-zero surface current leading to,

\be
X_{V} = \frac{\epsilon\, U}{2}\,\Big(\bar h''(\bar \zeta)\, X^{\bar \zeta }_{(0)}+h''( \zeta)\, X^{\zeta}_{(0)}\Big).\ee

Again, it is evident from (\ref{restf}) that there is a distortion of the particles after encountering the impulsive gravitational wave and the distortions are determined by superrotation parameter. Due to the presence of surface current, this distortion moves the test particles off the spatial 2-dimensional (as $X_V$ is non-zero) surface where they were initially situated at rest.

\section{Memory Effect on Null geodesics}
 \subsection{Setup}
\begin{figure}
\centering
\begin{center}
\includegraphics[scale=0.60]{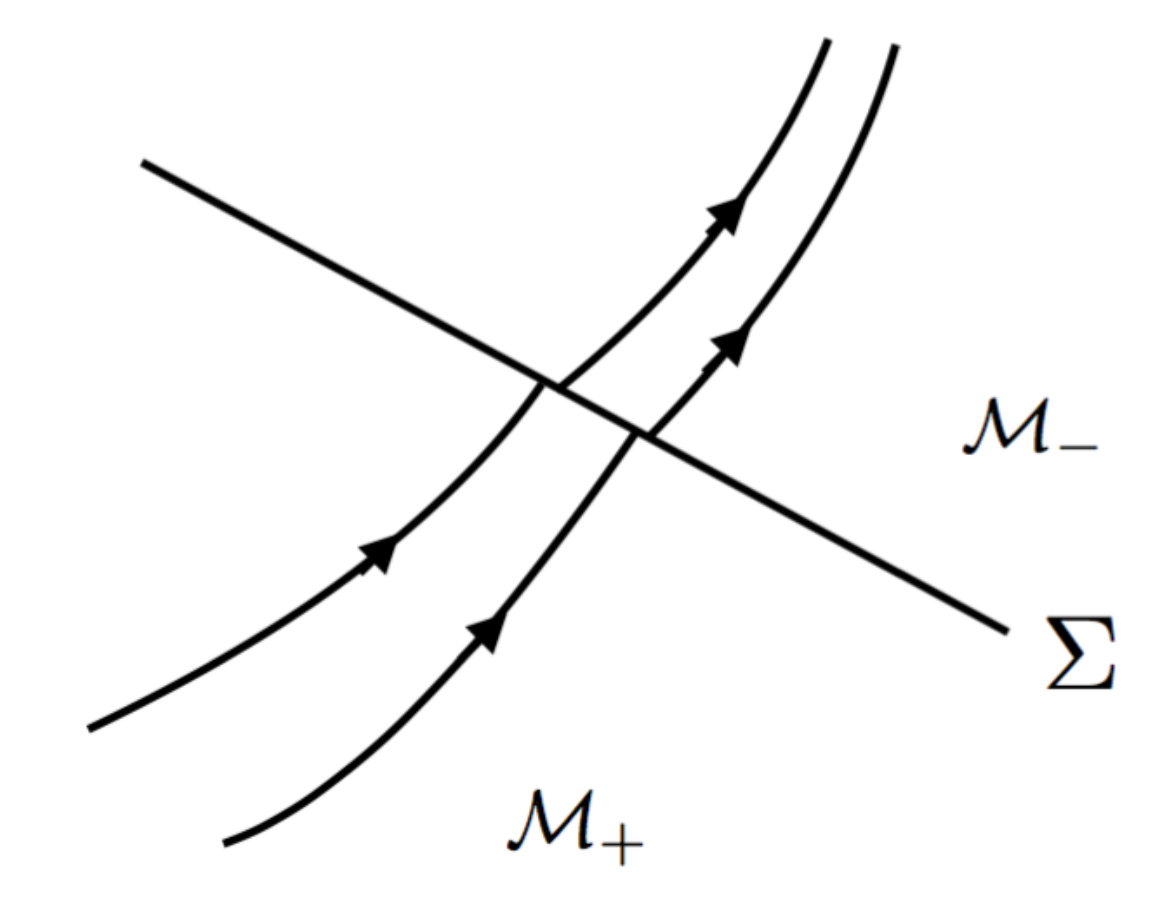}
\caption{Null geodesics crossing a null hypersurface $\Sigma$ from $\cal{M}_+$ to $\cal {M}_-$. Geodesics experience a jump upon crossing the hypersurface $\Sigma$ represented by soldering transformations. }
\label{ngc}
\end{center}
\end{figure}

Here, we consider a null congruence crossing a null hypersurface $\Sigma$ orthogonally, and study the change in the optical parameters like shear, expansion, etc. We compute these changes for a congruence defined by the tangent vector $N$ as it crosses the IGW. To do so, we first take a null congruence $N_+$  to the $`+'$ side of null-shell and evaluate the components of it in a common coordinate system installed across the shell. Next, the components of the same vector are evaluated to the $`-'$ side using the matching (soldering) conditions. In this procedure, the geodesic congruence suffers a jump and so as the expansion and shear corresponding to the congruence \cite{MO}. The setup is shown in the Fig.~(\ref{ngc}). \par
For null geodesics, $N^{\mu}$ is already defined in (\ref{auxn}), which can be thought as the tangent vector of some hypersurface orthogonal (locally with the null analogue of Gaussian normal coordinates) to null geodesics across $\Sigma.$ We denote $N_-$ to be the tangent to the congruence to the $-$ side of $\Sigma$. Mathematically the effect of the congruence $N_+$ crossing the null-shell is obtained by applying a coordinate transformation between the coordinates $x^{\mu}_+$ to a coordinate $x^{\mu}$ installed locally across the shell \cite{MO}\footnote{One could have done the transformation in the $'-'$ side equivalently.}. These relations between coordinates carry the soldering parameters producing memory effect. 

Let $N_0$ denotes the vector $N_+$ transformed in the common coordinate system and evaluated at the hypersurface $\Sigma.$. We take coordinates $x^{\mu}_-$ coinciding with $x^{\mu}$.  For expressing $N_0$ in a common coordinate system one needs the following transformation relation:
\begin{align}
N_{0}^{\alpha}(x) = \Big(\frac{\partial x_{+}^{\beta}}{\partial x^{\alpha}}\Big)^{-1}N_{+}^{\beta}\Big\vert_{\Sigma}.
\label{coordtt}\end{align}
The inverse Jacobian matrix $\Big(\frac{\partial x_{+}^{\beta}}{\partial x^{\alpha}}\Big)^{-1}$ is to be evaluated using the (``on-shell") soldering transformations \footnote{We call soldering freedom confined on the null hypersurface to be ``on-shell" see (\ref{on-shell}) of appendix~A. The  ``off-shell" version can be obtained from the  ``on-shell" version by extending the coordinate transformations off-the surface in the transverse direction, see (\ref{kerr4}) of appendix~A . }. The vector in the $'-'$ side is obtained by considering components of $N_0$ as initial values. Next, we compute the ``failure" tensor $B$ for the vector $N_0$ by taking the covariant derivative of it and projecting it on the hypersurface. This tensor encodes the amount of failure for the congruence to remain parallel to each other. 
\begin{align}
\begin{split}
B_{AB} = e^{\alpha}_{A}e^{\beta}_{B}\nabla_{\beta}N_{0\alpha}. 
\end{split}
\end{align}
Kinematical decomposition of this tensor leads to two measurable quantities (assuming zero twist for hypersurface orthogonal congruence):
\begin{align}
&\textrm{Expansion}:\,\, \Theta=\gamma^{AB}B_{AB},\\&
\textrm{Shear}:\,\, \Sigma_{AB}=B_{AB}-\frac{\Theta }{2}\gamma_{AB},
\end{align}
where $\gamma_{AB}$ is the induced metric on the null shell. Then memory of the IGW or shell is captured  through the jump of these quantities across $\Sigma$
\footnote{We discard the possibility of generating a non-vanishing twist to the congruence, initially possessing a zero-twist, after IGW pass through it. This is ensured from the evolution equation \cite{MO} of twist.}. Note that even if we use ``on-shell" version of the soldering transformations, jump in these quantities will be captured for a geodesic crossing the shell originating from $+$ side of the shell. 
\begin{align}
\begin{split} \label{jump}
&[\Theta]=\Theta  \big |_{\Sigma_+}-\Theta \big |_{\Sigma_-},\\&
[\Sigma_{AB}]=\Sigma_{AB} \big |_{\Sigma_+}-\Sigma_{AB}\big |_{\Sigma_-}.
\end{split}\end{align}
Next, we let the test geodesics to travel off-the shell, and we get additional contributions in jumps of the expansion and shear. This is achieved by transforming $B$ tensor further to the $'-'$ side of the shell via the following relation that is obvious from hypersurface orthogonal nature of the congruence:
\be
x^{\mu}_-\equiv x^{\mu}=x^{\mu}_0 + U N_0^{\mu}(x^{\mu})+\mathcal{O}(U^2).
\ee
 Note that all the additional contributions in expansion and shear are proportional to $U$, the parameter of off-shell extension, and would coincide with the jumps when we restrict the transformations on the shell, i.e. for $U=0$. After determining these off shell transformations we can evaluate the failure tensor. For determining $B$ tensor off the shell, we first compute $B_{AB} = e^{\alpha}_{A}e^{\beta}_{B}\nabla_{\beta}N_{0\alpha}$ and then pull it back to a point infinitesimally away from the shell. 
 \begin{align}\label{b-tensor}
\tilde{B}_{AB}(x^{\mu}) = \frac{\partial x_0^{M}}{\partial x^{A}}\frac{\partial x^{N}_0}{\partial x^{B}}B_{MN}(x^{\mu}_0).\end{align} 
In the far region of the shell, $B$ tensor is evaluated using $N_+=\lambda \partial_U$, and clearly it will be different from the expression we get from the above equation leading to the jumps in its different components. Next, we show these jumps for shells in Schwarzschild spacetime.

\subsection{ Memory for supertranslation like transformations} \label{result1}
We consider the case of the Schwarzschild black hole in Kruskal coordinate (\ref{smet}). For plane fronted waves, a similar kind of memory effect has been reported in \cite{MO}. Here we extend that for spherical waves. The details of off-shell transformations are given in appendix~A. We first compute the failure tensor off-the shell using the inverse Jacobian of transformations given in  
(\ref{Jacobian-1}),
\be
 N_{0}=\frac{e}{8 m^2} \Big(\partial_{U}+ \frac{1}{4F_{V}^2m^{2}e}\Big(F_{\theta}^{2}+\frac{F_{\phi}^{2}}{\sin^{2}\theta}\Big)\partial_{V}-\frac{F_{\theta}}{2 F_{V}m^{2}e }\partial_{\theta}-\frac{F_{\phi}}{2 F_{V} m^{2}e\sin^{2}\theta}\partial_{\phi}\Big)\Big\vert_{\Sigma}. 
\ee

With this and using (\ref{Jacobian4}) one can evaluate $\tilde{B}_{AB}$ off-the shell:
\begin{align}
\tilde{B}_{AB} =  \frac{1}{det(J)^{2}}\left(
\begin{array}{cc}
 \Big(1+\frac{2U}{e}T_{\phi\phi}\Big)^{2}+\frac{4U^{2}}{e^{2}}T_{\theta\phi}^{2} & -\frac{2U}{e}T_{\theta\phi}\Big(2+\frac{2U}{e}T_{\phi\phi}+\frac{2U}{e}T_{\theta\theta}\Big) \\
 -\frac{2U}{e}T_{\theta\phi}\Big(2+\frac{2U}{e}T_{\phi\phi}+\frac{2U}{e}T_{\theta\theta}\Big) &  \Big(1+\frac{2U}{e}T_{\theta\theta}\Big)^{2}+\frac{4U^{2}}{e^{2}}T_{\theta\phi}^{2}
\end{array}
\right) B_{MN},
\end{align}

The failure tensor $B_{AB}$ for the congruence with $N_+=\lambda \partial_U$\footnote{ $\lambda$ is arbitrary and only equal to $\frac{e}{8 m^2}$ on the shell} yields,
\begin{align}
B_{AB} =& e_{A}^{\alpha}e_{B}^{\beta}\bigtriangledown_{\beta}N_{\alpha} = - \Gamma^{V}_{BA}N_{V}.
\end{align}
 We find the $B$-tensor before and after the null congruence crosses the shell to be different which means there is a jump. Clearly for $U=0$, the off-shell failure tensor $\tilde{B}$ reduces to $B$ evaluated at the shell $\Sigma$. Now focusing particularly on supertranslation, and using (\ref{trans}) and (\ref{Jacobian-2}) one gets the following jumps across the shell: 
\begin{align}
 \begin{split}
&[\Theta ]=\frac{1}{(4 m^2)^2}\Big(T_{\theta\theta}+\csc(\theta)^2 T_{\phi\phi}+T_{\th}\, \cot(\theta)\Big),\\
&[\Sigma_{\theta\theta}]=\frac{1}{8 m^2}\Big(T_{\theta \theta }-T_{\phi \phi} \csc ^2(\theta )-T_{\th} \cot (\theta )\Big),\\&
[\Sigma_{\phi\phi}]=\frac{1}{ 8m^2}\Big(T_{\phi \phi }-T_{\theta \theta } \sin ^2(\theta )+\frac{T_{\th} \sin (2\theta )}{2}\Big),\\&
[\Sigma_{ \theta \phi}]=\frac{1}{(2m)^2}\Big(T_{\theta\phi}- T_{\phi} \cot (\theta )\Big).
 \end{split}
 \end{align}
 It is apparent from the above expressions that the jumps in different components of $B$ are related to supertranslation parameter $T(\theta,\phi)$, and we get a memory corresponding to this. We could have got jumps using the off-shell tensor $\tilde{B}$, in that case the jumps will differ from this by the terms proportional to $U.$ It should be noted, the jump in the shear term is arising due to the presence of IGW while in the expansion is due to the presence of shell stress tensor.

\subsection{ Memory for superrotation type transformations}

First we focus on the flat space which corresponds to $\eta=0,\Lambda=0$ for the metric mentioned in (\ref{constcurvature}). For this case using (\ref{trans1}) we have,
\be
 N_{0}= \Big(\partial_{U}-\frac{\epsilon\, \bar{h}^{''}(\bar \zeta)} {2\, V}\partial_{\zeta}-\frac{\epsilon \, h^{''}(\zeta)}{2V}\partial_{\zeta}\Big)+\mathcal{O}(\epsilon^2) \Big\vert_{\Sigma}.
\ee

Repeating the similar computation as in section (\ref{result1}),  and using (\ref{new6}) we get to leading order in $\epsilon$,
\begin{align}
\begin{split}
&[\Theta]=\mathcal{O}(\epsilon^2),\\&[\Sigma_{\zeta\zeta}] = \frac{\epsilon V h^{'''}(\zeta)}{2}+\mathcal{O}(\epsilon^2),
[\Sigma_{\bar{\zeta}\bar{\zeta}}]=\frac{\epsilon V \bar{h}^{'''}(\bar \zeta)}{2}+\mathcal{O}(\epsilon^2),\\&
[\Sigma_{\zeta\bar{\zeta}}]=\mathcal{O}(\epsilon^2),
[\Sigma_{\bar{\zeta}\zeta}] = \mathcal{O}(\epsilon^2).
\end{split}
\end{align}
Again it is evident that the jumps in these tensors capture the superrotation type of transformations.  We can perform the similar analysis for constant negative and positive curvature spaces which also described by the class defined in (\ref{constcurvature}).

\subsection{Memory for null-shell near a black hole horizon }
 In this case, the null surface is situated at $U=\epsilon$ of Schwarzschild black hole. Here, we use transformation relations mentioned in (\ref{trans2}). The inverse Jacobian and other details are displayed in appendix~B. Using (\ref{Jb3}) the tangent vector of past congruence can be written as,
 
\begin{align}
 N_{0} =\frac{e-2 V \epsilon }{8 m^2} \Big(\partial_{U}-B\partial_{\zeta}-\bar{B}\partial_{\bar{\zeta}}\Big)+\mathcal{O}(h^2,\bar h^2, \epsilon\, h, \epsilon\, \bar h)\Big\vert_{\Sigma}.
\end{align}
 
Performing similar analysis as done in section (\ref{result1})  we get,
\begin{align}
[\Theta]=& \frac{4 e^{2-m^2} \left((1+\zeta \bar \zeta) \left(h'(\zeta)+\bar h'(\bar \zeta)\right)-2 \bar \zeta h(\zeta)-2 \zeta \bar h (\bar \zeta)\right)}{1+\zeta \bar\zeta}+\mathcal{O}(h^2,\bar h^2, \epsilon\, h, \epsilon\, \bar h),\\
[\Sigma_{\zeta\zeta}] =&  -\frac{e}{4\,(1+\zeta\bar{\zeta})^{2}}\partial_{\zeta}\bar{B}+\mathcal{O}(h^2,\bar h^2, \epsilon\, h, \epsilon\, \bar h),
[\Sigma_{\bar{\zeta}\bar{\zeta}}] =-\frac{e}{4\,(1+\zeta\bar{\zeta})^{2}}\partial_{\bar\zeta}B+\mathcal{O}(h^2,\bar h^2, \epsilon\, h, \epsilon\, \bar h), \\
[\Sigma_{\zeta\bar{\zeta}}] =& \mathcal{O}(h^2,\bar h^2, \epsilon\, h, \epsilon\, \bar h),
[\Sigma_{\bar{\zeta}\zeta}] = \mathcal{O}(h^2,\bar h^2, \epsilon\, h, \epsilon\, \bar h).
\end{align}

$B$ and $\bar B$ are defined in (\ref{new4}).

\section{ Discussions}
In violent astrophysical phenomena, gravitational wave pulses with considerable strength are generated.  We have modelled such waves as null shells which represent history of such impulsive gravitational waves. The motivation of this work is to find the BMS memory effect near the vicinity of a black hole horizon. For this, we have reviewed the theory of null-shells endowed with BMS-like symmetries.  It is well known that these shells, in general, carry non-zero stress tensor along with IGW. We have studied the interaction of such IGW and stress tensor components on test particles. Let us list the findings of this note: 

\begin{itemize}

\item First, we have studied horizon-shell situated at the horizon of Schwarzschild black hole carrying IGW and evaluated its effect on the deviation vector connecting two nearby time like geodesics crossing the shell. In this case, where no surface current is present, the effect of IGW on test particles, initially at rest at the 2-dimensional spatial surface, is to displace them relative to each other within the surface. This displacement is characterized by a supertranslation parameter indicating the supertranslation memory effect. This work is an extension of earlier works on similar effects for planar gravitational waves \cite{BH-ijmp,Zhang1, Zhang2, Zhang3,  Stein}. 

\item Next, we considered the effect of superrotations on the deviation vector again on timelike congruence. We have exhibited the superrotation memory for Penrose's `spherical' IGW spacetime. In this case, due to the presence of non-zero surface current, the test particles initially situated at rest on the 2-dimensional spatial surface, will get displaced off the surface. A similar effect is expected to be exhibited by a null shell situated near to the event horizon of a black hole. This superrotation like memory effect near the horizon has not been considered earlier.

\item Futhermore, we turned our attention on null geodesics. We have shown the expansion and shear parameters of a null geodesic congruence encounter jumps characterized by supertranslation parameter upon crossing a null-shell situated at the Schwarzschild horizon. 

\item Finally, we have shown that a null-shell placed just outside of a black hole horizon exhibits superrotation like memory effect on the null observers in a similar fashion. We also indicated the similar effect for Penrose's `spherical' IGW spacetime.

\end{itemize}
The memory effect near the horizon of a black hole may offer many minute details of the symmetries appearing at the horizon region. We hope the more advanced versions of present day detectors might capture the effects considered here. A study of the near horizon analogue of (far region) memory effect is under progress and will be reported elsewhere.  Relation between BMS-like memories and the Penrose limit for exact impulsive wave spacetimes can offer interesting findings \cite{Shore:2018kmt}. As there exists some modified version of the BMS algebra on a null-surface situated at a finite location of a manifold \cite{Chandrasekaran:2018aop}, it will be interesting to investigate the role of those symmetries on the gravitational memory effect. The role of BMS-like symmetries in the study of quantum fluctuations of these spherical IGW background could be a fascinating area of research \cite{Hort1, hort2}.  It will also be interesting to see the implications of our results in the context of holography (AdS/CFT correspondence). Finally, to understand the connection between the quantum vacuum structure of gravity and these memories are going to be an active field of study. 

\vspace{-0.5cm}
\section*{Acknowledgements} \vspace{-0.2cm}
SB is supported by SERB-DST through the Early Career Research award grant ECR/2017/002124. Research of AB is supported by JSPS Grant-in-Aid for JSPS fellows (17F17023). 

\appendix 
\section{Off-shell extensions of soldering freedom on Killing horizons}
Here we provide necessary expressions useful for computation of stress tensor (\ref{ST}) on null-shell. 
The Scwarzschild Killing horizon case has been presented in \cite{Blau}, here we briefly review that.  Let us consider the case of Schwarzschild black hole mentioned in (\ref{smet}). The null shell is located at $U=0.$ On the $+$ side we can perform the following transformations,
\be
V_{+}= F(V, \theta,\phi), \theta_{+}=\theta, \phi= \phi_{+}.
\label{on-shell}\ee
We choose the intrinsic coordinates $x^{\mu}$ to be coordinates on the $-$ side. Introducing the following ansatz \cite{Blau},
\begin{align}
\begin{split} \label{kerr4}
V_{+}&=F(V,\theta,  \phi)+U A(V,\theta,\phi),U_{+}=U C(V,\theta, \phi),\\& \theta_{+}=\theta+U B^{\theta}(V,\theta,\phi), \phi_{+}= \phi+U B^{ \phi} (V,\theta, \phi),
\end{split}
\end{align}
and demanding the continuity of spacetime metric across the junction at leading order in $U$, we get,
 \begin{align}
\begin{split} \label{kerr5}
&C=\frac{1}{\partial_V F},\,A=\frac{e}{4}\partial_{V} F \Big((\frac{2}{e} \frac{\partial_{\theta}F}{\partial_V F})^2+\sin(\theta)^2(\frac{2}{e}\frac{1}{\sin(\theta)^2}\frac{\partial_{\tilde \phi}F}{\partial_{V} F})^2\Big),\\&
B^{\theta}=\frac{2}{e} \frac{\partial_{\theta}F}{\partial_V F},\,
B^{\phi}= \frac{2}{e}\frac{1}{\sin(\theta)^2}\frac{\partial_{ \phi}F}{\partial_{V} F}.
\end{split}
\end{align}
We then expand the tangential components of the metrics of both sides $\mathcal{M}_{+}$ and $\mathcal{M}_{-}$ to linear order in $U$ and read off the components $\gamma_{ab}$
using (\ref{jumpmet}),
$
\gamma_{ab}=N^{\alpha}[\partial_{\alpha} g_{ab}]= N^{U}[\partial_{U} g_{ab}]
$
with,
$
N^{U}=\frac{e}{8 m^2}.$
 We write down different components of $\gamma$ tensor:
\begin{align}
\begin{split} \label{kerr6}
&\gamma_{V a }=2\frac{\partial_{V}\partial_{a} F}{\partial_{V} F },\gamma_{\theta \theta }= 2\Big(\frac{\nabla_{\theta}^{(2)} \partial_{\theta} F}{\partial_{V} F }-\frac{1}{2}\Big(\frac{F}{\partial_{V} F}-V\Big)\Big),\\&\gamma_{\theta \phi }= 2\Big(\frac{\nabla_{\theta}^{(2)} \partial_{\phi} F}{\partial_{V} F  }\Big),\gamma_{\phi \phi }= 2\Big(\frac{\nabla_{\phi}^{(2)} \partial_{\phi} F}{\partial_{V} F }-\frac{1}{2}\sin(\theta)^2\Big(\frac{F}{\partial_{V} F}-V\Big)\Big).\end{split}
\end{align}
Using (\ref{kerr6}) and (\ref{junc8}), we can identify shell-intrinsic objects as follows,\begin{align}
\begin{split} \label{kerr7}
&p=-\frac{1}{16 \pi} \gamma_{VV}=-\frac{1}{8\pi}\frac{\partial_{V}^2 F}{\partial_{V} F},j^{A}=\frac{1}{32 m^2 \pi}\sigma^{AB}\frac{\partial_{B}\partial_{V}F}{\partial_{V} F},\\&\mu =-\frac{1}{32 m^2 \pi \partial_{V} F}\Big(\nabla^{(2)} F-F +V\partial_{V}F\Big).
\end{split}
\end{align}
For supertranslation we need to replace $F(V,\theta,\phi)$ by $V+ T(\theta,\phi).$ Then only $\mu$ is nonzero,
\be
\mu =-\frac{1}{32 m^2 \pi}\Big(\nabla^{(2)} T(\theta,\phi)-T(\theta,\phi) \Big).
\label{mu}\ee
\par
\vskip 0.2 cm

{\bf {\small{ Constant curvature spacetimes}}}
\vskip 0.2cm
Now we consider the class of metrics given by (\ref{constcurvature}) and extend the symmetry transformations mentioned in (\ref{trans1}). We start with the following ansatz,
\begin{align}
\begin{split} \label{new1}
&V_+=V(1-\frac{\epsilon\,\tilde \Omega(\zeta,\bar \zeta)}{2})+U  A(V,\zeta,\bar \zeta), U_+=U C(V,\zeta,\bar \zeta),\\&
\zeta_+=\zeta+\epsilon\, h(\zeta)+ U B(V,\zeta,\bar \zeta), \bar \zeta_+= \bar \zeta+ \epsilon\, \bar h(\bar \zeta)+U  \bar B(V,\zeta,\bar \zeta),
\end{split}
\end{align}
where $\tilde\Omega(\zeta,\bar\zeta)$ is defined in (\ref{Omega}) and we work upto linear order in $\epsilon.$ Again demanding the continuity of full spacetime metric across the junction at leading order in $U$ we get ,
\begin{align}
\begin{split} \label{new2}
&A=\frac{\epsilon\,\eta\,\tilde \Omega(\zeta,\bar\zeta)}{2},C=1+\frac{\epsilon\,\tilde \Omega(\zeta,\bar \zeta)}{2}, \\&
B=\frac{\epsilon  \left((1+\eta  \zeta \bar \zeta) \left(\bar h''(\bar \zeta) (1+\eta  \zeta \bar \zeta)-2 \eta  \zeta \bar h'(\bar \zeta )\right)+2 \eta ^2 \zeta^2 \bar h(\bar \zeta)-2 \eta  h(\zeta)\right)}{2 V},\\&
\bar B=\frac{\epsilon  \left((1+\eta  \zeta \bar \zeta) \left(h''(\zeta) (1+\eta  \zeta \bar \zeta)-2 \eta  \bar \zeta h'(\zeta)\right)+2 \eta ^2 \bar \zeta^2 h(\zeta)-2 \eta  \bar h(\bar \zeta)\right)}{2 V},
\end{split}
\end{align}
where we have used (\ref{Omega}). Now we can compute the $\gamma_{ab}$ as before. For simplicity, we write the results of flat space time i.e $\eta=0, \Lambda=0.$ The $\eta, \Lambda \neq 0$ case can be done analogously. So for the flat space we get,
\begin{align}
\begin{split}
\gamma_{V\zeta}=h''(\zeta), \gamma_{V\bar\zeta}=\bar h''(\bar \zeta), \gamma_{ \zeta \zeta}=V\,h'''(\zeta), \gamma_{ \bar \zeta \bar \zeta}=V\,\bar h'''(\bar\zeta).
\end{split}
\end{align}
Consequently, we will only have non-vanishing currents.
\be
j^{\zeta}=\frac{\epsilon\, \bar h''(\bar \zeta)}{16\pi\, V^2}, j^{\bar \zeta}=\frac{\epsilon\,  h''( \zeta)}{16\pi\, V^2}.
\ee

\subsection*{\small{Null surface near Killing Horizon} }
Lastly, we present the off-shell extension of the symmetry transformations mentioned in (\ref{trans2}). Keeping only those terms which are linear in $h(\zeta),\bar h(\bar \zeta)$ and $\epsilon$, we get, 
\begin{align}
\begin{split} \label{new3}
&U_+= (U-\epsilon) A, V_+= -\frac{a\,\tilde \Omega(\zeta,\bar \zeta)}{b\,\epsilon}+V(1-\tilde \Omega(\zeta,\bar \zeta))+(U-\epsilon) C, \\&
\zeta_+=\zeta+h(\zeta)+(U-\epsilon) B, \bar \zeta_+=\bar \zeta+\bar h(\bar \zeta)+(U-\epsilon) \bar B.
\end{split}
\end{align}
$a=4 m^2, b=-\frac{8 m^2}{e}$ and $\tilde \Omega(\zeta,\bar\zeta)$ is defined in (\ref{Omega}). Then demanding the continuity of the metric across the null shell upto $\mathcal{O}(U-\epsilon)$ yields,

\begin{align}
\begin{split} \label{new4}
&A=1+\frac{ \left((\zeta \bar \zeta+1) \left(h'(\zeta)+\bar h'(\bar \zeta)\right)-2 \bar \zeta h(\zeta)-2 \zeta \bar h(\bar \zeta)\right)}{1+\zeta \bar \zeta}, C=\mathcal{O}(h(\zeta)^2,\bar h(\bar \zeta)^2),\\&
B=-2 e^{1-m^2} \Big(2 h(\zeta)-(\zeta \bar \zeta+1) \left((\zeta \bar \zeta+1) \bar h''(\bar \zeta)-2 \zeta \bar h'(\bar \zeta)\right)-2 \zeta^2 \bar h(\bar \zeta)\Big),\\&
\bar B=-2 e^{1-m^2} \Big(2 \bar h(\bar \zeta)-(\zeta \bar \zeta+1) \left((\zeta \bar \zeta+1) h''(\zeta)-2 \bar \zeta h'(\zeta)\right)-2 \bar \zeta^2 h(\zeta)\Big).
\end{split}
\end{align}

\section{Jacobian of transformations}

Here we present the expression for {\textit{Jacobian} matrix mentioned in (\ref{coordtt}) which is crucial for computation of the memory effect through (\ref{jump}). We use the expressions  (\ref{kerr4}) and (\ref{kerr5}) to get
\begin{align}
\begin{split}
\Big(\frac{\partial x_{+}^{\beta}}{\partial x^{\alpha}}\Big)^{-1}\Big|_
{U=0} = \frac{1}{2m^2\,e}\left(
\begin{array}{cccc}
2 m^2 e F_{V} & 0 & 0 & 0 \\
 \frac{1}{2 F_{V}}\Big(F_{\theta}^{2}+\frac{F_{\phi}^{2}}{\sin^{2}\theta}\Big) & \frac{2 m^2\,e}{F_{V}} & -\frac{2 m^2\,e\,F_{\theta}}{F_{V}} & -\frac{2 m^2\, e\, F_{\phi}}{F_{V}} \\
 -F_{\theta} & 0 & 2 m^2\,e & 0 \\
 -\frac{F_{\phi}}{\sin^{2}\theta} & 0 & 0 & 2 m^2 e\\
\end{array}
\right).\end{split}
\label
{Jacobian-1}\end{align}
For the case of supertranslation, it further simplifies.
\begin{align}
\begin{split}
\Big(\frac{\partial x_{+}^{\beta}}{\partial x^{\alpha}}\Big)^{-1}\Big|_
{U=0} =\frac{1}{2m^2\,e}\left(
\begin{array}{cccc}
2 m^2 e  & 0 & 0 & 0 \\
 \frac{1}{2}\Big(T_{\theta}^{2}+\frac{T_{\phi}^{2}}{\sin^{2}\theta}\Big) & 2 m^2\,e & -2 m^2\,e\,T_{\theta} & -2 m^2\,e\, T_{\phi} \\
 -T_{\theta} & 0 & 2 m^2\,e & 0 \\
 -\frac{T_{\phi}}{\sin^{2}\theta} & 0 & 0 & 2 m^2 e\\
\end{array}
\right).\end{split}
\label{Jacobian-2}\end{align}

For off-shell, we already know that null congruence to the $-$ side is related to the coordinates on the surface via the following expression,
\begin{align*}
x^{\alpha} = x_{0}^{\alpha}+UN_{0}^{\alpha}
\end{align*}
The B-tensor in this mapping would give us the off shell extended version. The transformation can be written as,
\begin{align}
\tilde{B}_{AB} = \frac{\partial x_{0}^{M}}{\partial x^{A}}\frac{\partial x_{0}^{N}}{\partial x^{B}}B_{MN}
\end{align} 
$B_{MN}$ is nothing but the components of $B-tensor$ calculated on the shell. We would be considering here only BMS case. 

\begin{align*}
 \frac{\partial x_{0}^{M}}{\partial x^{A}} = \frac{\delta^{B}_{A}}{(\delta^{B}_{M}-U\frac{\partial N^{B}}{\partial x^{N}})}
=\frac{I}{J}. \end{align*}
Here $J=(\delta^{B}_{M}-U\frac{\partial N^{B}}{\partial x^{N}})$, and inverse of the $J$ matrix is then given by, 
\begin{align}
\Big(\delta^{B}_{M}-U\frac{\partial T^{B}}{\partial x^{N}}\Big)^{-1} = \frac{1}{det(J)}\left(
\begin{array}{cc}
 1+\frac{2U}{e}T_{\phi\phi} & -\frac{2U}{e}T_{\theta\phi} \\
 -\frac{2U}{e}T_{\theta\phi} &  1+\frac{2U}{e}T_{\theta\theta}
\end{array}
\right)
\label{Jacobian4}
,\end{align}
where determinant of $J$ is given by,
\begin{align*}
det(J) =1+\frac{2U}{e}(T_{\theta\theta}+F_{\phi\phi})-\frac{4U^{2}}{e^{2}}(T_{\theta\phi}^{2}-T_{\theta\theta}T_{\phi\phi}) .
\end{align*}
 
Next, using (\ref{new1}) and (\ref{new2}) for flat space we get,
\begin{align}
\begin{split} \label{new6}
\Big(\frac{\partial x_{+}^{\beta}}{\partial x^{\alpha}}\Big)^{-1}\Big|_
{U=0} =
\left(
\begin{array}{cccc}
  1-\frac{\epsilon}{2}(h^{'}(\zeta)+\bar{h}^{'}(\bar \zeta)) & 0 & 0 & 0 \\
0 & 1+\frac{\epsilon}{2}(h^{'}(\zeta)+\bar{h}^{'}(\bar \zeta)) & \frac{V\epsilon h^{''}(\zeta)}{2} & \frac{V\epsilon \bar{h}^{''}(\bar \zeta)}{2} \\
 -\frac{\epsilon \bar{h}^{''}(\bar \zeta)}{2\,V} & 0 & 1-\epsilon h^{'}(\zeta) & 0 \\
 -\frac{\epsilon h^{''}(\zeta)}{2V} & 0 & 0 & 1-\epsilon\bar{h}^{'}(\bar \zeta) \\
\end{array}
\right) +\mathcal{O}(\epsilon^2).
\end{split}
\end{align}

For the null surface near Killing horizon, using (\ref{new3}) and (\ref{new4}) we get ,
\begin{align}
\begin{split} \Big(\frac{\partial x_{+}^{\beta}}{\partial x^{\alpha}}\Big)^{-1}\Big|_
{U=\epsilon}=\left(
\begin{array}{cccc}
 1-\frac{ \left((\zeta \bar \zeta+1) \left(h'(\zeta)+\bar h'(\bar \zeta)\right)-2 \bar \zeta h(\zeta)-2 \zeta \bar h(\bar \zeta)\right)}{1+\zeta \bar \zeta} & 0 & 0 & 0 \\
 0 &1+\tilde \Omega (\zeta,\bar \zeta) & \frac{(a+b\, V\,  \epsilon ) \partial_{\zeta} \tilde \Omega(\zeta,\bar \zeta)}{b\, \epsilon } & \frac{(a+b\, V\, \epsilon ) \partial_{\bar \zeta}\tilde  \Omega(\zeta,\bar \zeta)}{b \, \epsilon } \\
 - B(\zeta,\bar \zeta) & 0 & 1- h'(\zeta) & 0 \\
 -\bar B(\zeta,\bar \zeta) & 0 & 0 & 1- \bar h'(\bar \zeta) \\
\end{array}
\right).
\end{split}
\label{Jb3}\end{align}

\bibliographystyle{JHEP}

\end{document}